\documentclass{nature}
\usepackage[letterpaper]{geometry}
\usepackage{color}
\usepackage{amsmath}
\usepackage{graphicx}
\usepackage{amssymb}
\usepackage{esint}
\long\def\symbolfootnote[#1]#2{\begingroup%
\def\thefootnote{\fnsymbol{footnote}}\footnote[#1]{#2}\endgroup}

\title{MD-Predicted Phase diagrams for Pattern Formation due to Ion Irradiation}

\author{Scott A. Norris$^{1,3}$, Juha Samela$^{2}$, Laura Bukonte$^{2}$, Marie Backman$^{2}$, Djurabekova Flyura$^{2}$, Kai Nordlund$^{2}$, Charbel S. Madi$^{3}$, Michael P. Brenner$^{3}$, \& Michael J. Aziz$^{3}$}

\begin{document}

\maketitle

\begin{affiliations}
 \item Southern Methodist University, Dallas, TX 75205
 \item 2 Department of Physics and Helsinki Institute of Physics, P.O. Box 43, FIN-0014 University of Helsinki, Helsinki, Finland
 \item Harvard School of Engineering and Applied Sciences, Cambridge, MA  02138
\end{affiliations}

\begin{abstract}
\let\thefootnote\relax\footnotetext{Submitted Sept. 15, 2010}

Energetic particle irradiation of solids can cause surface ultra-smoothening
\cite{yamada-etal-2001-MSER}, self-organized nanoscale pattern formation
\cite{chan-chason-JAP-2007}, or degradation of the structural integrity
of nuclear reactor components \cite{baldwin-doerner-NF-2008}. Periodic
patterns including high-aspect ratio quantum dots \cite{facsko-etal-SCIENCE-1999},
with occasional long-range order \cite{ziberi-etal-PRB-2005} and
characteristic spacing as small as 7 nm \cite{wei-etal-2008-CPL},
have stimulated interest in this method as a means of sub-lithographic
nanofabrication \cite{cuenat-etal-2005-AM}. Despite intensive
research there is little fundamental understanding of the mechanisms
governing the selection of smooth or patterned surfaces, and precisely
which physical effects cause observed transitions between different
regimes \cite{ziberi-etal-APL-2008,madi-etal-2008-PRL} has remained
a matter of speculation \cite{davidovitch-etal-PRB-2007}. Here we report the first
prediction of the mechanism governing the transition from corrugated surfaces
to flatness, using only parameter-free molecular dynamics simulations of single-ion impact
induced crater formation
as input into a multi-scale analysis, and showing
good agreement with experiment.
Our results overturn the paradigm attributing these phenomena
to the removal of target atoms via sputter erosion.  Instead, the
mechanism dominating both stability and instability is shown to be the impact-induced redistribution of
target atoms that are not sputtered away, with erosive effects being
essentially irrelevant. The predictions are relevant
in the context of tungsten plasma-facing fusion
reactor walls which, despite a sputter erosion rate that is essentially
zero, develop, under some conditions, a mysterious nanoscale topography leading
to surface degradation. Our results suggest that degradation
processes originating in impact-induced target atom redistribution effects
may be important, and hence
that an extremely low sputter erosion rate is an insufficient design
criterion for morphologically stable solid surfaces under energetic particle irradiation.
\end{abstract}

At irradiation energies between $10^{2}\,\mathrm{eV}-10^{4}\,\mathrm{eV}$,
the irradiation process is dominated by the nuclear collision cascade
caused by ion impact \cite{sigmund-PR-1969,sigmund-JMS-1973}. Displaced
atoms that reach the surface with enough kinetic energy to leave are
permanently sputtered away; all other displaced atoms come to rest
within the solid or on the surface after phonon emission times of
$\sim10^{-12}$ seconds. These processes contribute \emph{prompt}
\emph{erosive} \cite{bradley-harper-JVST-1988,makeev-etal-NIMB-2002}
and \emph{prompt redistributive} \cite{carter-vishnyakov-PRB-1996,moseler-etal-SCIENCE-2005,davidovitch-etal-PRB-2007}
components of morphology evolution, respectively, and are collectively
denoted $P\left[\mathbf{x}\right]$.  For most materials other than elemental metals,
the damage resulting from these
collisions quickly ($\sim10^{-3}$ seconds) leads to the amorphization
of a thin layer of target material.  Over much longer time scales
($\sim100$ seconds), mass transport by kinetic relaxation processes
causes a \emph{gradual} \emph{relaxational} effect \cite{bradley-harper-JVST-1988,umbach-etal-PRL-2001}.%
\symbolfootnote[2]{
Note that our focus on amorphous materials precludes potentially confounding effects due to singular
crystallographic energetics and kinetics, such as the Villain instability \cite{villain-1991:MBE},
in which surface diffusion is destabilizing and may be responsible for pattern formation on faceted single-crystal metal
surfaces at certain temperatures \cite{chan-chason-JAP-2007}.
}
Hence, to the prompt term $P\left[\mathbf{x}\right]$ we add a phenomenological
term for the gradual relaxation regime denoted $G\left[\mathbf{x}\right]$,
assuming a mechanism of ion-enhanced viscous flow, which is expected
to predominate in irradiated amorphous materials near room temperature \cite{umbach-etal-PRL-2001}.
The prompt and gradual contributions to the rate of motion of the surface
in the normal direction $v_{\mathbf{n}}$ superpose: \begin{equation}
v_{\mathbf{n}}=P\left[\mathbf{x}\right]+G\left[\mathbf{x}\right].\label{eqn: p-and-g}\end{equation}

The prompt regime may be characterized using molecular dynamics (MD)
simulations \cite{moseler-etal-SCIENCE-2005,kalyanasundaram-etal-APL-2008} or experimental
methods \cite{costantini-etal-PRL-2001}. Given data from many impact
events, we may obtain the ``crater function'' $\Delta h\left(\mathbf{x}-\mathbf{x}',\,\theta\right)$
describing the average change in local surface height at a point $\mathbf{x}$
resulting from a single-ion impact at $\mathbf{x}'$, with incidence
angle $\theta$.%
%!!!you need a different symbol-footnote because #2 is used earlier on the same page.
\symbolfootnote[3]{
In principle, $\Delta h$ also contains an explicit dependence
on the initial surface shape $h(\mathbf{x})$ (as opposed to the \emph{implicit}
dependence that arises via the angle-dependence).  This effect is described in
\cite{norris-etal-2009-JPCM}, but is very difficult to capture with molecular dynamics,
so we leave its analysis for future work.
}
We then upscale the crater function into a continuum partial
differential equation for the surface evolution using a multi-scale framework.
The theoretical formalism for this process is described elsewhere
\cite{norris-etal-2009-JPCM}; here we provide a brief summary of
the important points for the linear case. Given the crater function
$\Delta h$ and the flux distribution $I\left(\mathbf{x}\right)$,
we write the prompt contribution to surface evolution as a flux-weighted
integral of the crater function \cite{aziz-MfM-2006,davidovitch-etal-PRB-2007}:
\begin{equation}
P\left[\mathbf{x}\right]=\int I\left(\mathbf{x}'\right)\Delta h\left(\mathbf{x}-\mathbf{x}',\,\theta\right)\, d\mathbf{x}'\label{eqn: integral-formulation}\end{equation}
A well-known observation in the field is that scale of the craters
is much smaller than the scale of the resulting pattern and of the
flux distribution. To exploit this fact, we employ a formal multiple-scale
analysis, based on a small parameter $\varepsilon$ related to the
ratio of impact scales to pattern scales. This formalism allows ready
separation of the spatial dependence of the crater function (fast)
from that of the surface shape (slow), and leads eventually to an
upscaled description of the prompt regime of the form\begin{equation}
P\left[\mathbf{x}\right]=\left(IM^{\left(0\right)}\right)+\varepsilon\nabla_{\mathbf{S}}\cdot\left(IM^{\left(1\right)}\right)+\frac{1}{2}\varepsilon^{2}\nabla_{\mathbf{S}}\cdot\nabla_{\mathbf{S}}\cdot\left(IM^{\left(2\right)}\right)+\dots.\label{eqn: p-result}\end{equation}
Here the $\nabla_{\mathbf{S}}$ represent surface divergences, and
the (increasing-order) tensors $M^{\left(i\right)}$ are simply the
angle-dependent
\emph{moments} of the crater function $\Delta h$. This compact formulation
is interesting for two reasons.  First, the moments are readily obtainable
directly via MD simulation, and they converge with far fewer trials
than do descriptions of the entire crater function. Second, while atomistic
methods have been used in the past to obtain the amplitude of a single
term in a PDE obtained via phenomenological modeling \cite{enrique-bellon-PRL-2000,moseler-etal-SCIENCE-2005,zhou-etal-PRB-2008,headrick-zhou-JPCM-2009}, we believe this is the first derivation of an entire PDE from molecular
dynamics results.

Equation \eqref{eqn: p-result} describes the prompt regime $P\left[\mathbf{x}\right]$.
To fully capture the surface dynamics, we add to this
a relaxation mechanism $G\left[\mathbf{x}\right]$ associated with ion-enhanced viscous flow \cite{umbach-etal-PRL-2001}.
Together, $P\left[\mathbf{x}\right]$ and $G\left[\mathbf{x}\right]$
completely determine surface morphology evolution via Equation
\eqref{eqn: p-and-g}. From this (nonlinear) equation, pattern-forming
predictions are then obtained by examining stability of the the \emph{linearized}
equation as a function of the laboratory incidence angle $\theta$. The derivation
follows that in \cite{bradley-harper-JVST-1988}, and in an appropriate
frame of reference one finds that the magnitude of infinitesimal perturbations
$h$ away from a flat surface evolve, to leading order in $\varepsilon$, according
to the PDE \begin{equation}
\frac{\partial h}{\partial t}=\left(S_{X}\left(\theta\right)\frac{\partial^{2}h}{\partial x^{2}}+S_{Y}\left(\theta\right)\frac{\partial^{2}h}{\partial y^{2}}\right)-B\nabla^{4}h,\label{eqn: linear-stability}\end{equation}
where the angle-dependent coefficients
\begin{equation}
\begin{aligned}
S_{X}\left(\phi\right) & = I_0 \frac{d}{d\phi}\left[M^{\left( 1 \right)}\left(\phi\right)\cos\left(\phi\right)\right]\\
S_{Y}\left(\phi\right) & = I_0 M^{\left( 1 \right)}\left(\phi\right)\cos\left(\phi\right)\cot\left(\phi\right)
\end{aligned}
.\label{eqn: moment-to-linear-coeff}
\end{equation}
are determined from the \emph{first} moments obtained via MD,
and the constant coefficient $B$ is estimated from independent experiments.
The structure of Equation \eqref{eqn: linear-stability} indicates
that linear stability is determined strictly by the signs of the calculated
coefficients $\left(S_{X},S_{Y}\right)$: for values of $\theta$
where either of these coefficients is negative, linearly unstable modes exist
and we expect patterns, whereas for values of $\theta$ where they are
both positive we expect flat, stable surfaces.

Existing uses of MD crater data for investigations of surface pattern-forming
are entirely numerical in nature \cite{kalyanasundaram-etal-JPCM-2009},
and could be viewed as a scheme for \emph{numerically} integrating
Equation \ref{eqn: integral-formulation}.  In contrast, our \emph{analytical}
upscaling of Equation \ref{eqn: integral-formulation} illustrates
exactly which qualities of the crater -- namely, its moments -- play
the dominant role in surface evolution.  Furthermore, this analytical
form can be linearized, allowing predictions of stability boundaries,
and changes to those boundaries as crater shape is varied. A crucial
component of our approach is that the crater function $\Delta h$
-- and hence the moments $M^{\left(i\right)}$ -- contains the contributions
of both erosion and mass redistribution. Whereas these effects have
traditionally been treated separately via unrelated phenomenological
models, viewing the crater function as fundamental integrates erosion
and redistribution into a unified description, allowing both processes
to be treated identically and readily separated and compared. Indeed,
this approach has permitted us to confirm for the first time conjectures
\cite{aziz-MfM-2006,davidovitch-etal-PRB-2007,kalyanasundaram-etal-APL-2008}
that the stability of irradiated surfaces could be dominated by redistributive
effects, with erosion -- long assumed to be the source of roughening
-- being essentially irrelevant.

We obtain angle-dependent moments from a series of MD simulations, in an
environment consisting of an amorphous, 20x20x10 nm Si target consisting
of 219,488 atoms created using the
Wooten/Winer/Weaire (WWW) method \cite{wooten-winer-weaire-1985-PRL},
and then annealed with the EDIP potential \cite{bazant-kaxiras-justo-1997-PRB}.
This gives an optimized amorphous structure where most of the Si atoms
have coordination number 4. The target was then bombarded with Ar
at 100 eV and 250 eV. During bombardment, the interaction between Si atoms was
again described using the EDIP potential, which gives a good agreement
between simulated and experimental sputtering yields \cite{samela-etal-2007-NIMB},
whereas the Ar-Si interaction was a potential calculated for the Ar-Si
dimer \cite{nordlund-etal-NIMB-1997}. The kinetic energy was gradually
removed during the simulations from the 1 nm borders of the substrate
to prevent it re-entering the impact area via the periodic boundary
conditions used in the simulations. The ambient temperature in the
simulations was 0 K. The simulation arrangements and their suitability
for cluster and ion bombardment simulations are discussed in more
detail in the supplement and in Refs. \cite{nordlund-etal-1998-PRB,ghaly-nordlund-averback-1999-PMA,samela-etal-2005-NIMB,samela-etal-2007-NIMB}.

% ------------------ FIGURE 1 --------------------------

For each energy, two hundred impacts were simulated at
each incidence angle in 5-degree increments
between 0 and 90 degrees, yielding moments as summarized in Figure 1.
From the initial and final atomic positions, moments were obtained
using the method described in \cite{norris-etal-2009-JPCM}.
Briefly, by assuming that densities in the amorphous layer attain
a steady state (i.e., that defect distributions immediately project
to the target surface), we obtain erosive moments by assigning a height
loss at each location proportional to the number of sputtered atoms
originating from that location, while redistributive moments were
obtained by assigning height losses at initial atomic positions and
height gains at final atomic positions. For use within our analytical
framework, we also fit the moments to Fourier series constrained by
symmetry conditions and by the observation that all moments tend to
zero at $\theta=\pm90^{\circ}$.
For both energies, the redistributive first moments are much larger
in magnitude than -- and have the opposite sign of -- the erosive moments.
The implication is that redistributive effects completely dominate erosive effects,
except possibly at the highest (grazing) angles where all moments tend to zero.

% ------------------ FIGURE 2 --------------------------
%

To corroborate this finding, we calculate in Fig. 2 the coefficients in equation \eqref{eqn: moment-to-linear-coeff}
for the 250 eV moments, and compare the pattern wavelengths they predict to experimental
observations in the same environmental conditions \cite{madi-etal-JPCM-2009} (clean linear
experimental data at 100 eV are not currently available).  The agreement is
generally good at intermediate angles, but several discrepancies between theory
and experiment should be addressed.
First, the small quantitative difference between predicted and observed bifurcation
angles -- which depends only on the shape of $S_{X}\left(\theta\right)$ -- could readily arise from the
approximate nature of the classical potential, on which our simulations are based
(unlike the bifurcation angle, precise wavelength values depend on the value of $B$,
which could only be estimated).  Second, the measured moments do not
predict a transition to perpendicular modes at the highest angles;
this could be due to our neglect of explicit curvature-dependence in the crater function,
but additional physical effects such as shadowing and surface channeling
-- not addressed here -- are known to be important at grazing angles.
Third, we find no prediction via MD of
the experimentally-observed perpendicular-mode ripples at low angles\cite{madi-etal-2008-PRL};
%This phenomenon is poorly understood; while Kalyanasundaram et al
%\cite{kalyanasundaram-etal-JPCM-2009} report ripples in the parametric
%vicinity of this regime with their numerical approach, the ripples
%are of the wrong orientation.
indeed, our redistribution-dominated continuum
PDE is \emph{maximally stable} at low angles, and equations of the
general form \ref{eqn: linear-stability} are anyway unable to generate
the constant-wavelength or ``Type I'' bifurcation that is observed
\cite{davidovitch-etal-PRB-2007}. These observations, together
with the experimental observation that low-angle ripples develop over much
longer timescales than their high-angle counterparts \cite{madi-etal-JPCM-2009},
suggest that low-angle perpendicular-mode ripples are not due to crater-function
effects at all. It has already been observed \cite{davidovitch-etal-PRB-2007}
that the low-angle Type I bifurcation is consistent with any of several
\emph{non-local} physical mechanisms such as \emph{gradual} stress buildup and relaxation
\cite{davidovitch-etal-PRB-2007}, or non-local damping \cite{facsko-etal-PRB-2004}.
None of these effects would be captured in the (prompt, local) crater function,
and this observation motivates future studies incorporating such physics.

Despite the limitations of our approach, when one considers the lack of any free
parameters in the theory, the agreement for the diverging-wavelength or ``Type II''
bifurcation \cite{cross-hohenberg-RoMP-1993} near 50 degrees is remarkably good.
The agreement remains good even when the erosive coefficients are omitted,
and the similar shapes of the redistributive moments at 100 and 250 eV is
consistent with the reported
\cite{madi-etal-JPCM-2009} energy-insensitivity of the stability boundary.
The most striking aspect of this result is its logical conclusion that
erosion effects are essentially irrelevant for determining the patterns:
according to Fig. 2 the contributions of redistributive
effects to the $S$ coefficients, which determine stability and patterns,
are about an order of magnitude greater and opposite in sign.
This conclusion overturns the erosion-based paradigm that has dominated
the field for two decades \cite{bradley-harper-JVST-1988} and we
suggest its replacement with a redistribution-based paradigm. An important
direction for future research is identifying the energy range over
which this conclusion holds. Although it is conceivable that erosive
effects might become non-negligible at higher ion energies, or for
dense metals where heat spikes enhance sputtering yields,
it remains to be determined whether erosion is actually important for stability
or pattern formation in any physical experiment to date.  Preliminary
analysis of simulations at 1 keV are consistent with our conclusions from the
results at 100 eV and 250 eV, and experimental results at 1 keV lead to the
same conclusion \cite{madi-etal-unpub-2010}.

Non-erosive ion impact-induced atom redistribution at surfaces has
heretofore not been considered in the design of plasma-facing fusion
reactor wall materials, where low sputter yield has been an important
design criterion in the selection of tungsten for stable surfaces
that must be exposed to large plasma particle fluxes for extended
periods. Because the average helium ion energy is only $\sim60\,\mathrm{eV}$
and the threshold energy for sputter removal of tungsten is $\sim100\,\mathrm{eV}$,
this material has been considered impervious to the effects of erosion.
Within the erosion-based paradigm of pattern formation, the nanoscopic surface
morphology that evolves on tungsten surfaces under some such conditions has
therefore appeared mysterious \cite{baldwin-doerner-NF-2008}.
However, because the negligibility of erosive effects does not prevent
redistributive effects from causing pattern-forming instabilities, as we have
shown here through a crater-function analysis, atom redistributive effects may be
important contributors to the origin of these mysterious morphologies, and we present
in the Supplement a re-analysis of data gathered in \cite{henriksson-etal-NIMB-2006}
that supports this idea.  If this conjecture turns out to be correct, then ultimately
crater function engineering considerations may provide a more refined materials design
criterion than simply a low sputter yield.

%\changed{However, simulations indicate
%that although He irradiation does not lead to the erosion of W, it does
%cause the formation of nanoscale He bubbles .
%The rupture of these bubbles, in addition to the direct displacive processes
%discussed above, appear to contribute to a net downstream displacement over many
%impacts (see Appendix).

%% Here is the endmatter stuff: Supplementary Info, etc.
%% Use \item's to separate, default label is "Acknowledgements"

\begin{addendum}
 \item S.A.N. and M.P.B. were supported by the
National Science Foundation through the Division of Mathematical Sciences,
M.P.B. was additionally supported through the Harvard MRSEC and the
Kavli Insitute for Bionano Science and Technology at Harvard University.
M.J.A. and C.S.M. were supported by Department of Energy grant DE-FG02-06
ER46335. The work of J.S., L.B., M.B., D.F., and K.N. was performed within the Finnish
Centre of Excellence in Computational Molecular Science (CMS), financed
by The Academy of Finland and the University of Helsinki; grants of
computer time from the Center for Scientific Computing in Espoo, Finland,
are gratefully acknowledged. We also thank M.J. Baldwin, N. Kalyanasundaram,
and H.T. Johnson for helpful discussions.
 \item[Competing Interests] The authors declare that they have no
competing financial interests.
 \item[Correspondence] Correspondence and requests for materials
should be addressed to M.J.A.~(email: aziz@seas.harvard.edu).
\end{addendum}

\clearpage

\begin{figure}
\begin{centering}
\includegraphics[width=5.5in]{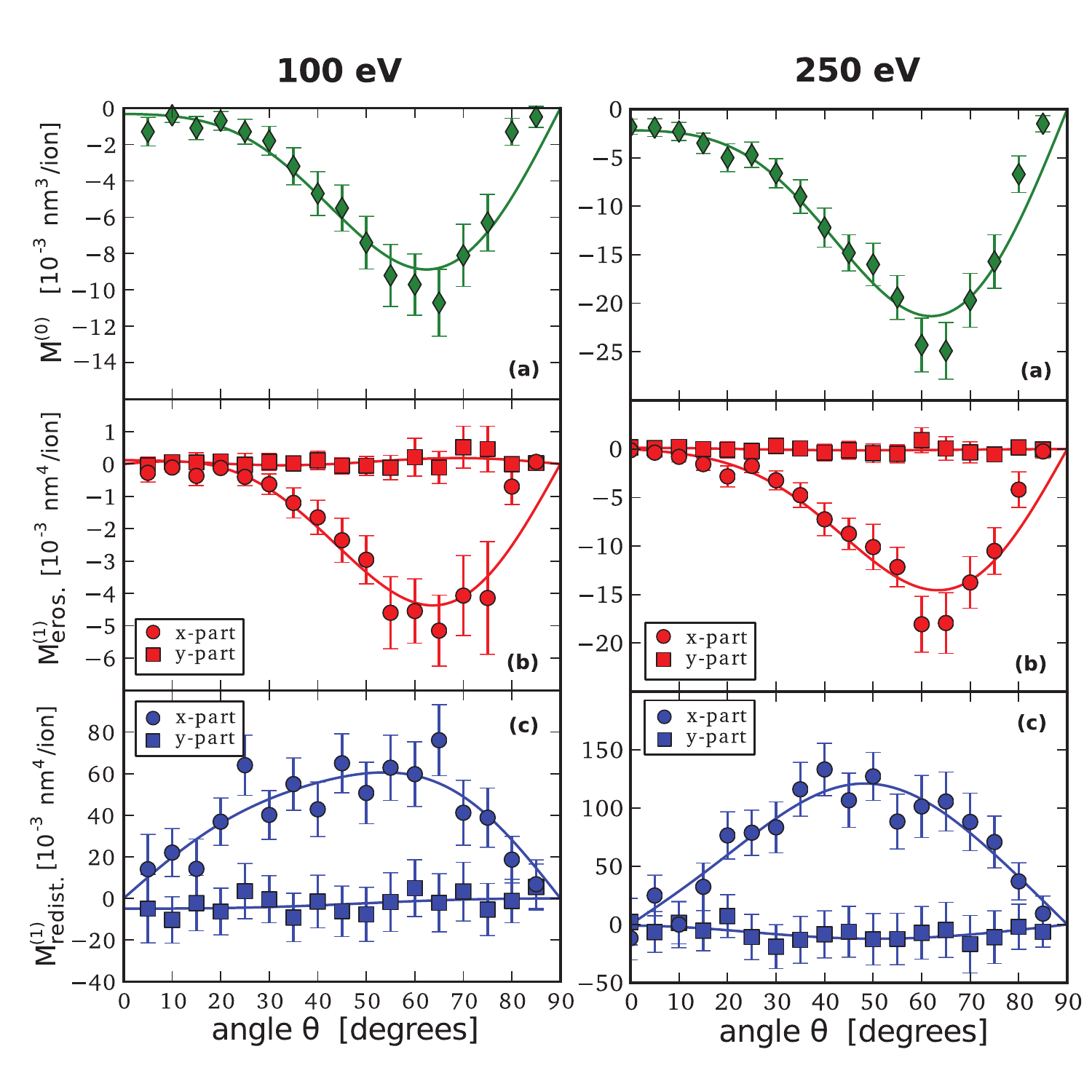}
\par\end{centering}

\caption{Fitted angle-dependent moments of the crater function $\Delta h\left(x,y\right)$
as determined from molecular dynamics, for both 100 eV and 250 eV.  (a) Zeroeth erosive
moment $M^{\left(0\right)}$ (sputter yield times atomic volume).
(b,c) First erosive (b) and redistributive (c) moments $M^{\left(1\right)}$.
Each of the latter contain components in both the $x$ (downbeam) and $y$
(crossbeam) directions, with the latter expected to be zero from symmetry
arguments \cite{norris-etal-2009-JPCM}.  At both energies, redistribution
dominates erosion.}

\label{fig: moments}
\end{figure}

\begin{figure}
\begin{centering}
\includegraphics[width=5in]{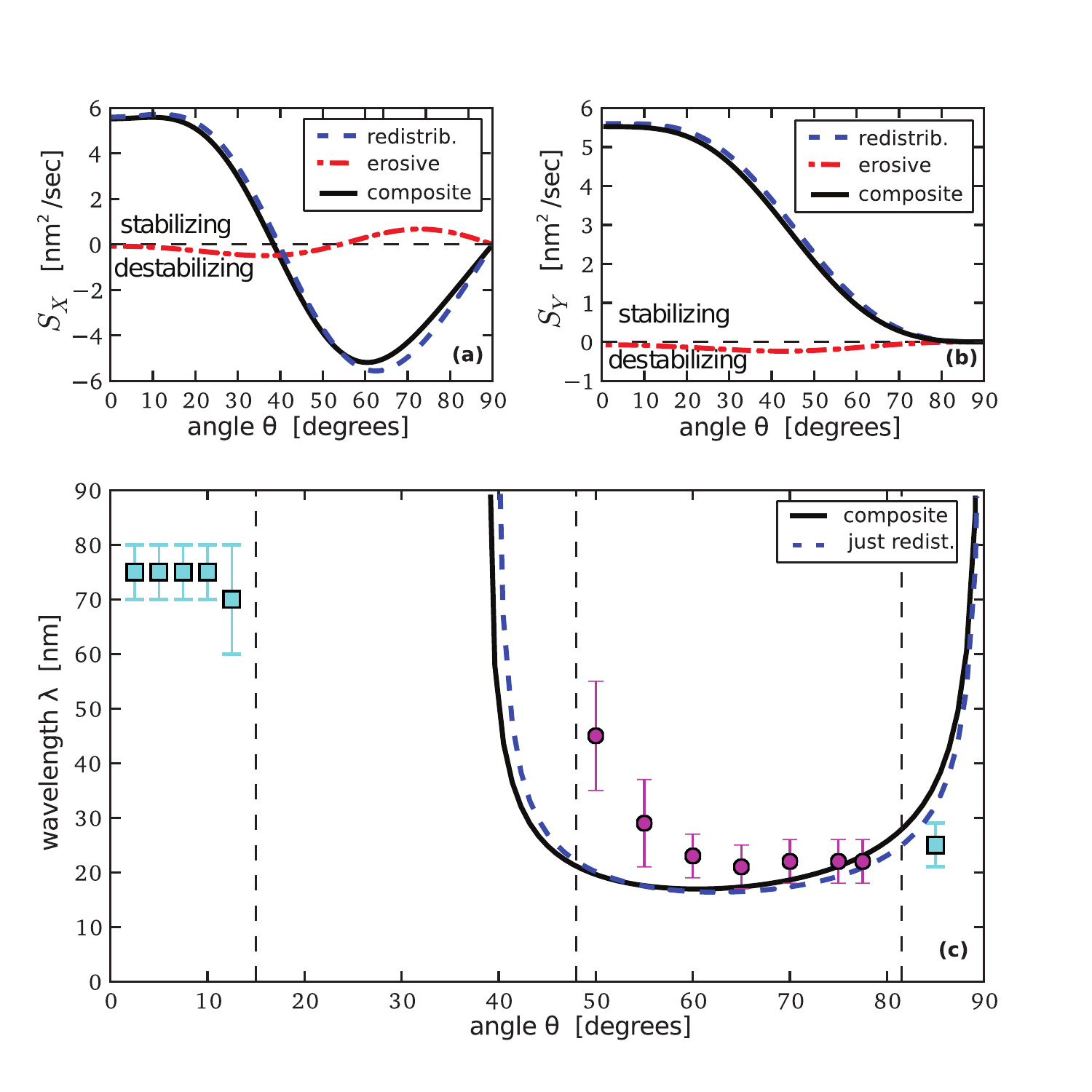}
\par\end{centering}

\caption{ Comparison between predicted and measured wavelength for 250 eV
Ar$\to$ Si.  Dimensional coefficients were calculated with a
flux of $f=3.5\times10^{15}\,\mathrm{ions}/\left(\mathrm{cm}^{2}\mathrm{s}\right)$
for comparison with experiment \cite{madi-etal-JPCM-2009}.
(a,b) Coefficients of $\frac{\partial^{2}h}{\partial x^{2}}$
and $\frac{\partial^{2}h}{\partial y^{2}}$ in the linearized evolution
equation \ref{eqn: linear-stability} using the experimental flux.
These coefficients are dominated by redistributive effects.
(c) Comparison of predicted ripple wavelengths with average experimentally-observed
wavelengths. Circles / squares indicate experimental patterns with
wavevector parallel / perpendicular to the
ion beam, and the vertical dashed black lines indicate experimental
phase boundaries. On top of this, the solid line indicates our
predicted wavelengths (which are all parallel-mode), and the dashed
blue line indicates the wavelengths predicted if erosion were neglected
entirely.}

\label{fig: predicted-wavelengths}
\end{figure}

\clearpage

\begin{appendix}

\section{Simulation Methods}

In the simulations, the actual impact induced structural deformations
can be detected only in a well relaxed silicon structure. A stress
existing within the material as a result of incomplete relaxation
of the structure before the simulation can induce displacements of
silicon atoms upon impact. For example, our test simulations showed
that a structure created by annealing silicon in MD is not dense enough
to model the real amorphous silicon. The surface of such a structure
collapses upon impact. It is possible to relax the internal stresses
by bombarding the surface with silicon atoms before the actual simulation.
However, the relaxation will be not uniform. Therefore, we have used
the Wooten/Winer/Weaire (WWW) method \cite{wooten-winer-weaire-1985-PRL}.

The WWW method is a computer algorithm that generates realistic random
network models of amorphous silicon. In this method, the structure
is described with positions of N atoms and a list of bonds between
the atoms. After a random switch of two bonds, the structure is relaxed
using an interatomic potential \cite{mousseau-barkema-JPCM-2004,alfthan-kuronen-kaski-2003-PRB}.
In connection to the structural relaxation, the Berendsen pressure
control algorithm is used to relax the diagonal components of the
pressure tensor to zero \cite{berendsen-etal-JCP-1984}. The algorithm
is computationally demanding and therefore it is possible to fully
optimize only relatively small structures of a few thousand atoms.
Therefore, the optimized block must be copied and the copies joined
to create an optimized silicon structure large enough for impact simulations.
Our tests showed that the best approach is to partially optimize a
rather large (10x10x10 nm) building block instead of building the
structure of very small fully optimized blocks. The latter approach
can induce artificial internal shear stresses in spite of the optimization
(note that pressure control at periodic boundaries an an orthogonal
cell does not necessarily remove the nondiagonal, shear components
of the stress tensor). The optimization of a 10x10x10 nm amorphous
silicon structure was achieved using a parallelized implementation
of the WWW algorithm.

After the WWW optimization phase, the structure and the surface were
relaxed in MD using the EDIP potential, before it was used in the
actual impact simulations. The structure used in the simulations was
built of four identical optimized blocks. The density was $2.5\,\mathrm{g}/\mathrm{cm}^{3}$,
which indicates that the structure is dense and not likely to collapse
upon impact. Two-thirds of the silicon atoms had four neighbours,
the others had five neighbours, which is also a sign of a dense structure.
In a perfect amorphous network, all silicon atoms should have four
neighbors. However, the test with structures built of smaller blocks
that were better optimized than the 10x10x10 nm structure showed that
the behavior of the moments as a function of the impact angle are
very similar as those reported in the main article.

To simulate ion irradiation of bulk silicon samples, we have used
periodic boundary conditions with slab boundary conditions (free boundary
in the incoming ion direction, and periodic boundaries in the other
two lateral directions). The bottom 1 nm layer of atoms in the simulation
cell were held fixed, and temperature scaling was also applied to
the atoms in a 1 nm thick layer above it. A border region of 1 nm
thickness in the lateral directions was cooled during the simulations.
The role of the cooling zones is to prevent shock waves and phonons
to re-enter the impact area via the periodic boundaries. The size
of the simulation box (20x20x10 nm) was chosen to be large enough
to contain not only the area containing the crater but also the area of small deformations
which reach 5-7 nm from the impact point.

In addition to the good amorphous silicon structure and cooling of
the boundaries, the quality of the repulsive Ar-Si potential affects
the outcome of impact simulations. The kinetic energy of the impacting
Ar atom is deposited first in relatively strong Ar-Si collisions where
the atoms come close each other. The repulsive Ar-Si potential used
in this study is calculated using density-functional-theory calculations
utilizing a numerical basis sets. This method is shown \cite{nordlund-etal-NIMB-1997}
to give a more accurate repulsive potential than the standard universal
ZBL potential \cite{ziegler-biersack-littmark-1985-SRIM}. The same
method was used to create Si-Si short-range repulsive potential which
was smoothly joined to the EDIP potential. This ensures that the possible
collisions between high-energy primary knock-on silicon atoms are
correctly modelled.

The same initial structure was used for all simulations. However,
to reduce as much as possible the effect of the initial surface structure
on the measured crater statistics, the following steps were taken.
First, the impact point was varied randomly over the entire surface,
with (periodic) cooling regions dynamically re-assigned for each simulation
so as to be maximally distant from the impact point. Second, the azimuthal
angle was varied randomly, and only the atoms within 7.5 nm of the
impact point were included in the moment calculations (i.e., cooling
zones and corners of the square box were ignored). Last, the global
average -- or ``background'' -- displacement (over
all simulations at a given impact angle) was subtracted from each
individual displacement before analysis. Because we varied the impact
point and azimuthal angle randomly, this global average ought to be
zero for a perfect target over enough trials, and subtracting it helps
to remove target-specific effects from our measurements. With this
setup, 200 simulations at each impact angle were performed,
which was sufficient to isolate the background displacement, and to
cancel the effect of thermal vibrations which were present in the
structure after impact.

A final challenge in analyzing the data arose due to a combination of
the amorphous nature of the target with the periodic boundary conditions.
On an ideal, very large MD target, the effect of the ion impact would
only be felt within a finite distance from the impact point.  However, to
allow the gathering of data within reasonable time, a limited
box size must be used, and the periodic box described above -- with cooling zones --
appears to be the most physically plausible way of accomplishing this.
However, a periodic box always means that one is truly simulating an infinite
number of simultaneous side-by-side impacts, and with a small enough domain, these impacts
can generate enough co-ordinated momentum transfer to shear the entire target,
especially for an amorphous target.  Indeed, within our target, the average downbeam
displacement of atoms was consistently a linear function of distance from the target's
rigid floor, except near the surface, where larger displacements were concentrated.

\begin{figure}
\begin{centering}
\includegraphics[width=5in]{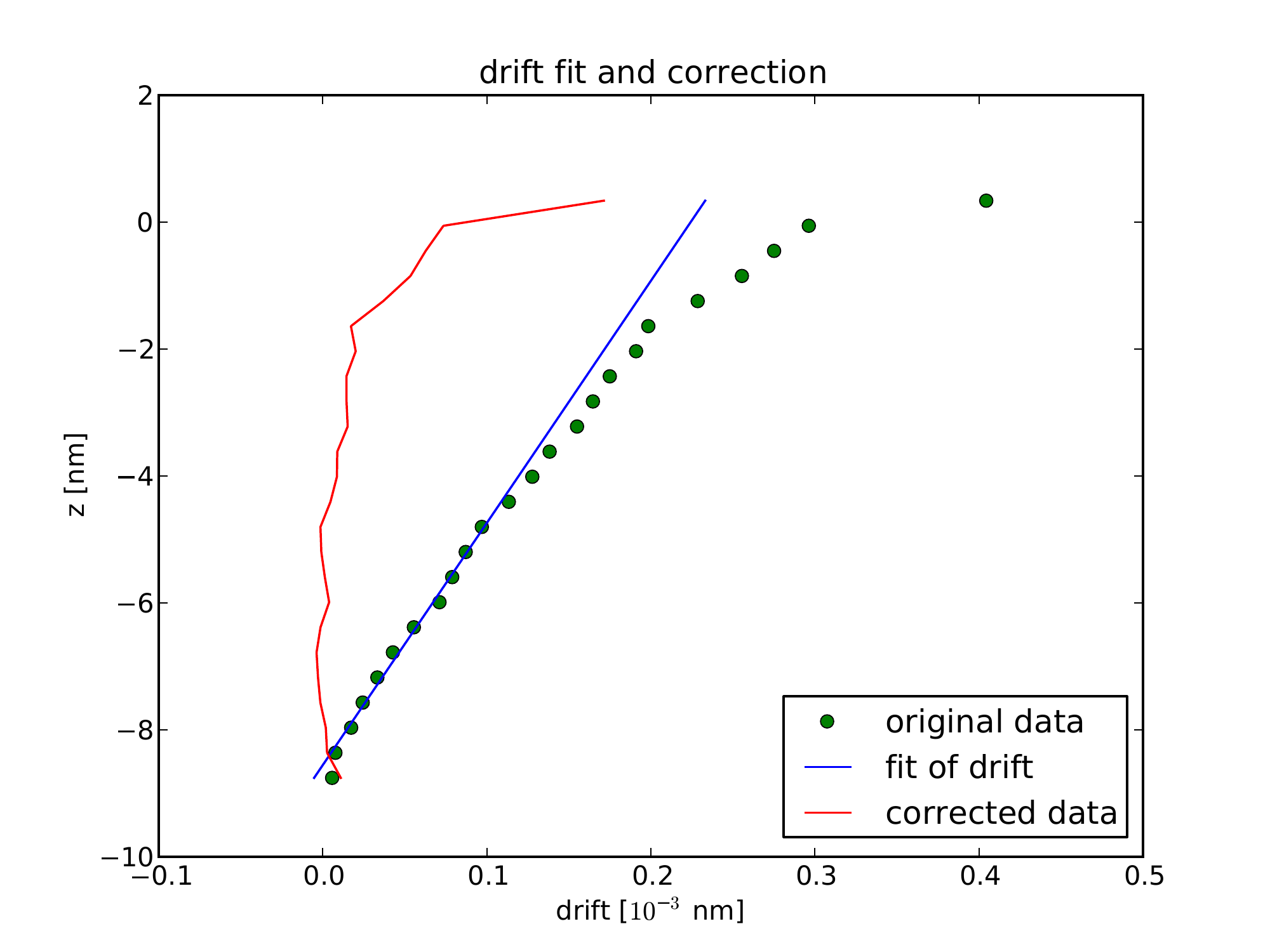}
\par\end{centering}

\caption{Illustration of the removal of shear at 60 degrees for irradiation at 250 eV.
Green dots are the original measurements, and the blue line represents a linear fit to
the bottom half of the dots, extrapolated to the surface.  The red line is the result
of subtracting this extrapolation from the original data, which localizes the displacements
to the vicinity of the surface.  All data are associated with a 2nm/9nm annulus that
masks the impact zone and the cooling boundary zone.}

\label{fig: shear-removal}
\end{figure}

To combat the problem of "global shear," we measured the average downbeam displacement
within different annular slices of target parallel to the surface, using inner/outer
radii of 2nm/9nm (i.e., away from both the impact and the cooling boundaries).  We then
fit the \emph{bottom} half of the resulting depth-dependent profile to a line using least
squares, and finally subtracted the \emph{extrapolation to the surface} of this fit from the overall displacement
field, as illustrated in Figure \ref{fig: shear-removal}.  The results localize the
displacements to within a few nanometers of the surface, which is consistent with measured
amorphous layer thicknesses of 3 nm for Si irradiated by Ar at 250 eV \cite{madi-etal-JPCM-2009}.
In the future, we will explore the response of a larger, hybrid target consisting of a 3 nm
layer of amorphous Si atop a crystalline (but not rigid) base.  However, we believe our
existing measurements are within the accuracy level of the other estimates in the paper.

\section{Moment Capture and Fitting}

Here we describe in more detail how we obtain moments, how we fit
them, and how we obtain the final linearized evolution equation (4).
For each simulated ion impact with our initially flat MD target, we
define a co-ordinate system $\left(x,y,z\right)$ centered at the
impact point with $z$ pointing normal to the surface, $x$ pointed
in the direction of the projected ion path, and $y$ perpendicular
to both $x$ and $z$ so as to complete a right-handed co-ordinate
system. Hence, the ion is always arriving from the \emph{negative}
\emph{$x$}-direction. Hence, the crater function $\Delta h\left(x,y;\,\theta\right)$
describes the height change associated with an impact at the origin,
of an ion with indicence angle of $\theta$ from the vertical, coming
from the negative $x$-direction.

After impact, we extract the moments \begin{equation}
\begin{aligned}M^{\left(0\right)}\left(\theta\right) & =\iint\Delta h\left(x,y\right)\,\mathrm{d}x\,\mathrm{d}y\\
M_{x}^{\left(1\right)}\left(\theta\right) & =\iint\Delta h\left(x,y\right)x\,\mathrm{d}x\,\mathrm{d}y\\
M_{y}^{\left(1\right)}\left(\theta\right) & =\iint\Delta h\left(x,y\right)y\,\mathrm{d}x\,\mathrm{d}y\end{aligned}
\end{equation}
using the method described in \cite{norris-etal-2009-JPCM}. Here,
it is important to note that the first moments $M^{\left(1\right)}$
contain both erosive and redistributive components (because the zeroeth
moment $M^{\left(0\right)}$ describes mass loss, and because redistribution
is mass-conserving, $M^{\left(0\right)}$ has no redistributive component).

Simulation sets were performed at 5-degree increments, and the average
of the resulting moments were fit to Fourier functions of the form

\begin{equation}
\begin{aligned}M_{x-odd} & =\sum_{n=1}^{3}a_{n}\sin\left(2n\theta\right)\\
M_{x-even} & =\sum_{n=1}^{3}b_{n}\cos\left(\left(2n-1\right)\theta\right)\end{aligned}
\label{eqn: fittings}\end{equation}
These fittings reflect the observation that all moments tend to zero
at $\theta=90^{\circ}$, and also their symmetries about $\theta=0$
(because a positive theta indicates an ion beam coming from the negative
$x$-direction, moments that are odd/even in $x$ should also be odd/even
in $\theta$). As seen in the main text, this method does not produce
perfect fits, but the use of simple Fourier modes eliminates potential
model-bias, while the restriction to a small number of terms excludes
inter-angle noise from the fitted curve.

For our data, the moment fits were given by:\begin{equation}
\begin{aligned}M^{\left(0\right)} & \approx-12.0\cos\left(\theta\right)+13.2\cos\left(3\theta\right)-3.46\cos\left(5\theta\right) & \left[10^{-3}\,\mathrm{nm}^{3}/\mathrm{ion}\right]\\
M_{x,\text{erosive}}^{\left(1\right)} & \approx-10.4\sin\left(2\theta\right)+5.81\sin\left(4\theta\right)-0.516\sin\left(6\theta\right) & \left[10^{-3}\,\mathrm{nm}^{4}/\mathrm{ion}\right]\\
M_{x,\text{redist.}}^{\left(1\right)} & \approx291\sin\left(2\theta\right)-39.6\sin\left(4\theta\right)-2.16\sin\left(6\theta\right) & \left[10^{-3}\,\mathrm{nm}^{4}/\mathrm{ion}\right]\end{aligned}
.\end{equation}
The moments $M_{y}^{\left(1\right)}$ are zero to within sampling
error, as expected from symmetry considerations.

\section{Analysis: from Moments to Coefficients }

For the general reduction of moments to (nonlinear) PDE terms $P\left[\mathbf{x}\right]$,
we refer to the framework derived in \cite{norris-etal-2009-JPCM}.
However, in the linear case discussed here, it is sufficient to consider
the linearization of the first-order term obtained by combining Equations
(1) and (3) from the main text: \begin{equation}
v_{\mathbf{n}}^{P}\left(\mathbf{x}\right)\approx\varepsilon\nabla_{\mathbf{S}}\cdot\left(IM^{\left(1\right)}\right)
\end{equation}
(in the linearization, the zeroeth- and second-moment terms do not contribute
to stability). Now, $\nabla_{S}$ indicates a \emph{surface} divergence,
and indeed this calculation is most naturally performed in a \emph{local}
co-ordinate system associated with the surface normal and projected
beam direction. In particular, both the flux $I\left(\phi\right)$
and the moments $M^{\left(i\right)}\left(\phi\right)$ depend on the
local incidence angle $\phi$, while the vector $M^{\left(1\right)}\left(\phi\right)$
is observed to always point in the direction $\mathbf{e}_{P}$ of
the projected ion beam.

Following \cite{norris-etal-2009-JPCM}, surface velocities at an
arbitrary point $\mathbf{x}$ will be calculated in a local co-ordinate
system $\left(U,V,W\right)$ centered at $\mathbf{x}$, where $\mathbf{e}_{W}=\mathbf{n}$
corresponds to the surface normal, $\mathbf{e}_{U}=\mathbf{e}_{P}$
corresponds to the \emph{downbeam }direction associated with the projected
ion beam, and $\mathbf{e}_{V}=\mathbf{e}_{W}\times\mathbf{e}_{U}$.
In this system the surface is described locally by the equation $W=H\left(U,V\right)$,
the ion flux is $I=I_{0}\cos\left(\phi\right)$, and the first moment
is $M^{\left(1\right)}=f\left(\phi\right)\mathbf{e}_{P}$, where $f\left(\phi\right)=M_{X}^{\left(1\right)}\left(\theta\right)$
as measured from MD. Now, as described in \cite{norris-etal-2009-JPCM},
it is sufficient for the purposes of calculating one surface divergence
to approximate $H\left(U,V\right)$ via\begin{equation}
H\approx\frac{1}{2}\left(H_{UU}U^{2}+2H_{UV}UV+H_{VV}V^{2}\right).\end{equation}
where $H_{UU}$, $H_{UV}$, and $H_{VV}$ describe the surface curvature
at $\mathbf{x}$. All other variable quantities can then be approximated
\emph{in the vicinity of $\mathbf{x}$} to first order in$\left(U,V\right)$
via: \begin{equation}
\begin{aligned}\mathbf{n} & \approx\left\langle -\frac{\partial H}{\partial U},\,-\frac{\partial H}{\partial V},\,1\right\rangle \\
\mathbf{e}_{P} & \approx\left\langle 1,\,-\cot\left(\phi_{0}\right)\frac{\partial H}{\partial V},\,\frac{\partial H}{\partial U}\right\rangle \\
\cos\left(\phi\right) & \approx\cos\left(\phi_{0}\right)+\frac{\partial H}{\partial U}\sin\left(\phi_{0}\right)\end{aligned}
\end{equation}
When we now take the surface divergence $\nabla_{S}=\left(\partial_{U},\,\partial_{V}\right)$
and evaluate at $\left(U,V\right)=\mathbf{0}$ (i.e., at $\mathbf{x}$),
we obtain in the local co-ordinate system\begin{equation}
v_{\mathbf{n}}^{P}\left(\mathbf{x}\right)\approx\varepsilon\nabla_{S}\cdot\left(IM^{\left(1\right)}\right)=\varepsilon I_{0}\left[S_{U}\left(\phi\right)H_{UU}+S_{V}\left(\phi\right)H_{VV}\right]\label{eqn: local-coords-evolution}\end{equation}
where\begin{equation}
\begin{aligned}S_{U}\left(\phi\right) & =\frac{d}{d\phi}\left[f\left(\phi\right)\cos\left(\phi\right)\right]\\
S_{V}\left(\phi\right) & =f\left(\phi\right)\cos\left(\phi\right)\cot\left(\phi\right)\end{aligned}
.\label{eqn: moment-to-linear-coeff2}\end{equation}
While linear in the local co-ordinates, Equation \eqref{eqn: local-coords-evolution}
is in general nonlinear in the lab fram. However, for stability studies
we need only the linearization of \eqref{eqn: local-coords-evolution}
in the lab frame, which in dimensional form is simply\begin{equation}
\left.\frac{\partial h}{\partial t}\right|_{\mathsf{prompt}}= I_{0}\left(S_{X}\left(\theta\right)\frac{\partial^{2}h}{\partial x^{2}}+S_{Y}\left(\theta\right)\frac{\partial^{2}h}{\partial y^{2}}\right)\label{eqn: linearization}\end{equation}
because, to linear order, \begin{equation}
\begin{aligned}\frac{\partial^{2}h}{\partial x^{2}} & \approx H_{UU}\\
\frac{\partial^{2}h}{\partial y^{2}} & \approx H_{VV}\\
S_{X}\left(\theta\right) & \approx S_{U}\left(\phi\right)\\
S_{Y}\left(\theta\right) & \approx S_{V}\left(\phi\right)\end{aligned}
\end{equation}
To expression \eqref{eqn: linearization} for the prompt regime we
add the linearization of the gradual regime associated with ion enhanced
viscous flow \cite{umbach-etal-PRL-2001}, which is a lubrication
approximation with the form \begin{equation}
\left.\frac{\partial h}{\partial t}\right|_{\mathsf{gradual}}=-B\nabla^{4}h.\end{equation}
Adding the prompt and gradual regimes, we obtain the evolution equation
(4) in the main text.

For completeness, we conclude with the functional forms of the coefficients
$\left(S_{X},S_{Y}\right)$ associatd with our fittings, which are
obtained from equations \eqref{eqn: moment-to-linear-coeff2}:\begin{equation}
\begin{aligned}S_{X}^{\mathsf{eros.}} & =-52.1\cos\left(\theta\right)-69.2\cos\left(3\theta\right)+132\cos\left(5\theta\right)-18.1\cos\left(7\theta\right)\\
S_{Y}^{\mathsf{eros.}} & =-50.5\cos\left(\theta\right)+24.7\cos\left(3\theta\right)+21.3\cos\left(5\theta\right)-2.58\cos\left(7\theta\right)\\
S_{X}^{\mathsf{redist.}} & =1450\cos\left(\theta\right)+3760\cos\left(3\theta\right)-1040\cos\left(5\theta\right)-75.6\cos\left(7\theta\right)\\
S_{Y}^{\mathsf{redist.}} & =3520\cos\left(\theta\right)+815\cos\left(3\theta\right)-231\cos\left(5\theta\right)-10.8\cos\left(7\theta\right)\end{aligned}
.\end{equation}

\section{Estimation of Viscous Flow coefficient}

The materials parameter $B$ appearing in Equation (4) of the main
text is defined \cite{orchard-ASR-1962} as\begin{equation}
B=\frac{\gamma d^{3}}{3\eta}\end{equation}
where $\gamma$ is the surface free energy, $d$ is the thickness
of the thin amorphous layer that is being stimulated by the ion irradiation,
and $\eta$ is the layer's viscosity. We assume the surface free energy
of amorphous silicon under ion irradiation to be equal to its value
in the absence of irradiation; the value of $\gamma=1.36\,\mathrm{J}/\mathrm{m}^{2}$
measured via molecular dynamics simulations by Vauth and Mayr \cite{vauth-mayr-PRB-2007}
happens to be numerically equal to that measured experimentally for
single-crystal Si(001) \cite{eaglesham-etal-PRL-1993}. For the amorphous
layer thickness, we directly measured $d\approx3.0\,\mathrm{nm}$
via cross-sectional transmission electron microscopy on samples irradiated
at normal incidence and 30 degrees from normal. Finally, we estimate
the viscosity of the top amorphous Si layer during irradiation to
be $\eta\approx6.2\times10^{8}\,\mathrm{Pa\, sec}$, as shown below.

The reciprocal of viscosity is the fluidity $\phi$, which is generally
understood to scale with the flux $f$ , and can be expressed in the
form\begin{equation}
\phi=H\times N_{\mathrm{DPAPS}}\end{equation}
where $H$ is the radiation-induced fluidity, and $N_{\mathrm{DPAPS}}$
is the average number of displacements per atom per second. Using
molecular dynamics simulations, Vauth and Mayr \cite{vauth-mayr-PRB-2007}
report $H=1.04\times10^{-9}\,\left(\mathrm{Pa\, dpa}\right)^{-1}$
at an energy $E=1\,\mathrm{keV}$ and temperature $T=300\mathrm{K}$
-- we use this value for our comparison with experiment, with the
caveats discussed below. The average number of displacements per atom
per second is given by \begin{equation}
N_{\mathrm{DPAPS}}=\frac{\Omega f}{d}N_{\mathrm{recoils}},\end{equation}
where $\Omega=.02\,\mathrm{nm}^{3}/\mathrm{atom}$ is the atomic volume
of silicon, $f=3.5\times10^{15}\,\mathrm{ions}/\left(\mathrm{cm}^{2}\mathrm{s}\right)$
is the experimental flux in the plane perpendicular to the ion beam,
$d\approx3\,\mathrm{nm}$ is the amorphous layer thickness, and $N_{\mathrm{recoils}}$
is the number of recoils generated per ion impact. To estimate $N_{\mathrm{recoils}}$,
we use the Kinchin-Pease \cite{gnaser-BOOK-1999} model for the gross
number of Frenkel pairs per incident ion, obtaining $N_{\mathrm{recoils}}=0.8E/E_{D}=6.7$,
where $E=250\,\mathrm{eV}$ is the ion beam energy and $E_{D}=15\,\mathrm{eV}$
is the displacement threshold energy of Si \cite{wallner-etal-JNM-1988}.
Taking all of these quantities, we obtain a value of the viscosity
of\begin{equation}
\eta=\frac{1}{H\times N_{\mathrm{DPAPS}}}=6.2\times10^{8}\mathrm{Pa\, sec}.\label{eqn: eta-estimate}\end{equation}

As discussed above, the value for $\eta$ listed in \eqref{eqn: eta-estimate}
is associated with Vauth and Mayr's value of $H=1.04\times10^{-9}\,\left(\mathrm{Pa\, dpa}\right)^{-1}$
at an energy $E=1\,\mathrm{keV}$ and temperature $T=300\mathrm{K}$.
In contrast, our experiments were carried out at $E=250\,\mathrm{eV}$,
and the irradiated sample is observed to reach temperatures of approximately
$450\,\mathrm{K}$. Hence, there is some uncertainty in our value
of $\eta$, which translates to uncertainty regarding the vertical
position of the theoretical curve in Figure 2 of the main text. For
the temperature difference, Vauth and Mayr observe $H$ to increase
weakly with increasing substrate temperature, which would shift the
curve upward; however, the shift would likely be less than a factor
of two. As for the energy difference, in a study of CuTi, another
amorphous material, Mayr et al \cite{mayr-etal-PRL-2003} found $H$
to either increase or decrease with recoil energy depending on the
details of the simulations; hence the theoretical curve in Fig. 2
of the main text could shift either upward or downward for MD simulations
at 250 eV, with a potential magnitude of perhaps a factor of two.

\section{Preliminary Supportive Data}
We believe the phenomena reported in this paper are applicable
to a wide variety of systems.  However, the main data sets leading to the
discovery of this result both describe low-energy irradiation of amorphous
silicon.  Therefore, we provide here some preliminary results associated with
higher-energy argon irradiation of silicon, and also for the case of helium-irradiated
tungsten which was referred to in our conclusion.

For silicon irradiated by argon at 1 keV, a much larger target must be used, with
dimensions of 40 x 40 x 10 nm, which makes simulations expensive.  However, we have
performed 30 simulations at 5-degree increments between 30 and 65 degrees, which spans
the experimental transition, and the results so far are consistent with those at 100
and 250 eV:  the redistributive contributions to the first moment dominate the erosive
contributions.

%\begin{figure}
%\begin{centering}
%\includegraphics[width=5in]{figures/moments-1keV}
%\par\end{centering}
%\caption{Moments calculated at a partial suite of angles for 1 keV.  These
%data are much noisier than those in the main text, but to within the error bars
%both the $\theta$-dependence, and the relative magnitudes, of the erosive and
%redistributive moments are similar to the results at 100 and 250 eV.}
%\label{fig: 1keV-moments}
%\end{figure}

For tungsten irradiated by helium at 100 eV, \emph{average} displacements are
so small that we were not able to isolate a clear signal against background noise
after 200 simulations.  However, using existing data from earlier work \cite{henriksson-etal-NIMB-2006},
we were able to calculate the \emph{cumulative} displacement field at a single angle
after 4,000 simulations.  In all of these simulations, not a single tungsten atom
was sputtered, yet a downbeam bias in the displacement field is clearly observed in
Figure 3. Hence a crater-function approach may enable better engineering of surface
stability under conditions where the sputter yield is zero but impact-induced target
atom displacements still occur.

\clearpage

\begin{figure}
\begin{centering}
\includegraphics[width=5in]{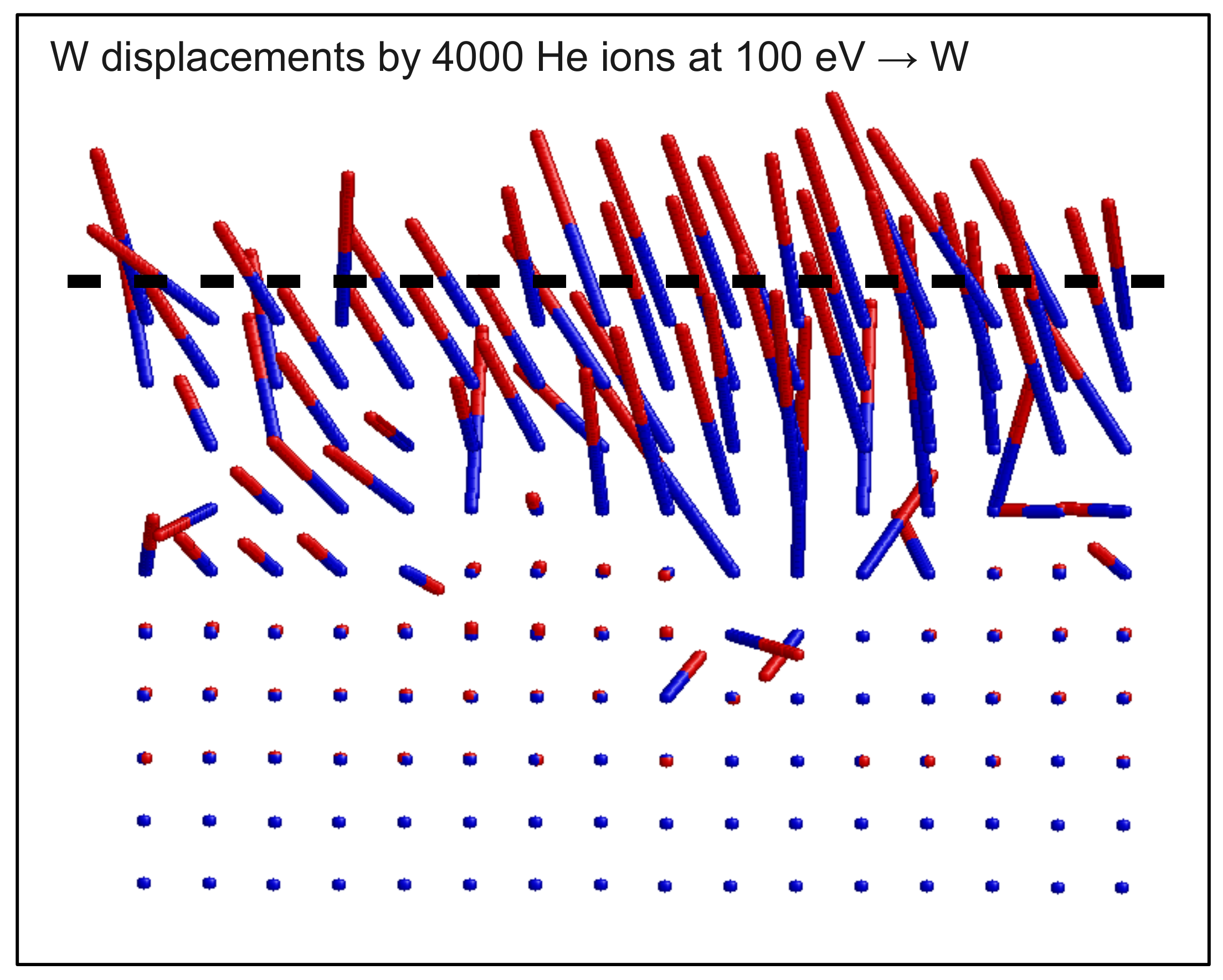}
\par\end{centering}

\caption{Cumulative displacement field for 100 eV He $\to$ W after 4,000 impacts.
Impinging helium ion comes from upper right, at an angle of 25 degrees from normal.
The sticks combine the initial and final position of
atoms that were initially in a half a unit cell thick cross section
through the simulation cell. The displacements are analyzed for
data initially simulated in Ref. \cite{henriksson-etal-NIMB-2006}. The blue ends
of the sticks indicate the initial and the red ends the final
positions of the atoms. The surface is located at the top.
No W atoms were sputtered, but a clear downbeam bias in the displacement field is seen.
}

\label{fig: cumulative-w}
\end{figure}

\end{appendix}

\clearpage

\bibliography{tagged-bibliography}

%%
%% TABLES
%%
%% If there are any tables, put them here.
%%

\end{document}